\documentclass[aps,prb,showpacs,reprint, superscriptaddress]{revtex4-1}

\usepackage{graphicx}
\usepackage{color}
\usepackage{amsmath}
\usepackage{bbm}
\usepackage{amssymb}
 
\usepackage[T1]{fontenc}

\input epsf

\begin{document}

\title[Universal properties of high-temperature superconductors from real-space pairing...]{Universal properties of high-temperature superconductors from real-space pairing III: The role of correlated hopping and intersite Coulomb interaction within the t-J-U model}


\author{Micha{\l} Zegrodnik}
\email{michal.zegrodnik@agh.edu.pl}
\affiliation{Academic Centre for Materials and Nanotechnology, AGH University of Science and Technology, Al. Mickiewicza 30, 30-059 Krak\'ow,
Poland}

\author{J{\'o}zef Spa{\l}ek}
\email{jozef.spalek@uj.edu.pl}
\affiliation{Marian Smoluchowski Institute of Physics, 
Jagiellonian University, ul. \L ojasiewicza 11,
30-348 Krak\'ow, Poland}

\begin{abstract}
We study the effect of the correlated hopping term and the intersite Coulomb interaction term on principal features of the $d$-$wave$ superconducting (SC) state, in both the electron and hole doped regimes within the t-J-U model. In our analysis we use the approach based on the \textit{diagrammatic expansion of the Gutzwiller wave function} (DE-GWF) which allows us to go beyond the renormalized mean field theory (RMFT). We show that the correlated hopping term enhances the pairing at the electron-doped side of the phase diagram. Moreover, the so-called non-BCS regime (which manifests itself by the negative kinetic energy gain at the transition to the SC phase) is narrowed down with the increasing magnitude of the correlated hopping $\sim K$. Also, the 
doping dependences of the nodal Fermi velocity and Fermi momentum, as well as the average number of double occupancies, are analyzed with reference to the experimental data for selected values of the parameter $K$. For the sake of completeness, the influence of the intersite Coulomb repulsion on the obtained results is provided. Additionally, selected results concerning the Hubbard-model case are also presented. A complete model with all two-site interactions is briefly discussed in the Appendix for reference.

\end{abstract}


\maketitle

\section{Introduction and motivation}\label{sec:intro}

Superconductivity (SC) in the strongly correlated electron systems has become one of the principal topics of research in the condensed matter physics, mainly because of the discovery of the high-temperature superconducting copper-based compounds. In this respect, the one-band Hubbard and t-J models have been extensively studied \cite{Anderson1997,Dagotto1990,Zhang1988,Ogata2008,Anderson1988,Anderson2004,Randeria2012} as it is believed that they capture the physics of the copper-oxygen planes which in turn are regarded as instrumental for achieving the SC state. Within the t-J model in which the local double occupancies are projected out from the wave function and the antiferromagnetic superexchange interaction appears, the SC phase emerges in a straightforward manner already at the renormalized mean-field theory level \cite{Anderson1988,Zhang1988_2,Edegger2007}. In the Hubbard model, in which the double occupancies are in general allowed, but significantly suppressed by the intraatomic Coulomb repulsion $\sim U$, more sophisticated methods are 
necessary to obtain the paired phase, e.g., the \textit{
variational Monte Carlo} (VMC) \cite{Eichenberger} or the \textit{diagrammatic expansion Gutzwiller wave function} (DE-GWF) \cite{Kaczmarczyk2013} approach.

A modified route is based on starting from the t-J-U model which combines the features appearing both in the Hubbard and in the t-J models. In this approach a relatively strong antiferromagnetic superexchange interaction is present among $3d$ Cu electrons and is induced by the virtual processes involving $2p$ bands of the neighboring $O^{2-}$ ions. Additionally, a small but nonzero number of double occupancies is allowed in the system at the same time. As it has been shown by us very recently \cite{Spalek2017} those two ingredients incorporated together into the correlated-electron picture beyond the renormalized mean field level (within the DE-GWF method), are essential in order to reproduce quantitatively selected principal experimental observations concerning the high temperature cuprate superconductors (HTS), at least within the full Gutzwiller wave-function solution. The physical justification of using such a model for the copper-oxides is provided in Ref. \onlinecite{Spalek2017}. A very good agreement 
with experiment has been obtained for relatively large values of the 
intrasite 
Coulomb repulsion ($U\approx 20|t|$, which leads to a very small admixture of the doubly 
occupied sites in the doped system) and for rather typical values of the electron hopping integrals (with nearest-neighbor amplitude $|t|\approx 0.35$ eV), as well as the exchange integral of the order of $J\approx0.3|t|$. In such a strongly correlated electron system the so-called correlated hopping term (often called the charge-bond interaction term), which has been disregarded in our previous considerations \cite{Spalek2017}, may also play some role. This term results from the off-diagonal element of the Coulomb interaction between the nearest neighboring lattice sites $\langle i,j\rangle$, with the corresponding two body integral $K_{ij}\equiv\langle\mathbf{i}\mathbf{i}|V(\mathbf{r}-\mathbf{r}')|\mathbf{i}\mathbf{j}\rangle$ (for definitions of those matrix elements see Appendix A). The importance of such a term has been pointed out previously \cite{Wojtowicz2001,Shavika,Stauber,Hubsch2006,Vitoriano2009,Gorski2011}. Also, it has been argued some time ago that for the case of the Hubbard $U$ strong 
suppression, the correlated hopping can lead to the paired phase already within the mean-field description \cite{Hirsh,Marsiglio,Micnas1990}. In the case of strongly correlated systems its influence on the SC phase has been investigated very recently \cite{Wysokinski2017} within the Hubbard model. 

In the series of papers \cite{Spalek2017,Zegrodnik2017} (Parts I and II, respectively) we have undertaken a systematic effort to compare our DE-GWF results for the t-J-U model in a quantitative manner with selected universal experimental characteristics for the cuprates. In this formulation the t-J and Hubbard models can be analyzed as limiting situations within a single theoretical framework. Here, we present the analysis of the $d$-$wave$ paired state in 
the 
presence 
of the correlated hopping within the t-J-U model and show our 
results in both the electron- and hole-doped regimes. We supplement our analysis also with the corresponding non-zero direct intersite Coulomb repulsion of magnitude $V_{ij}\equiv\langle\mathbf{i}\mathbf{j}|V(\mathbf{r}-\mathbf{r}')|\mathbf{i}\mathbf{j}\rangle$ which leads to the so-called t-J-U-V model, considered by us very recently in the context of charge-ordered-phase stability \cite{Abram2016}. In this manner, our model contains practically all relevant one- and two-site interactions within the single-band model \cite{Laughlin2014}, analyzed in the strong-correlation limit (cf. Appendix A). We focus on how the added terms influence the selected principal features of the high-$T_C$ superconductors which have been analyzed already within the t-J-U model\cite{Spalek2017}. We believe that such a question is important in order to see whether those features that have been claimed as universal \cite{Spalek2017} survive even in the case when the correlated-hopping and the intersite Coulomb-repulsion terms are 
included. In 
particular, we analyze the persistence of the BCS-like and non-BCS regimes \cite{Deutscher2005,Carbonne2006,Molegraaf2002,Gianetti2011}, the doping dependence of both the nodal Fermi velocity and the Fermi momentum \cite{Zhou2003,Kordyuk2005,Borisenko2006,Hashimoto2008}, as well as the critical concentrations for the SC dome in the phase diagram. It should be noted that experimental reports concerning these aspects are not so numerous when it comes to the electron-doped systems. In particular, the non-BCS regime has not been identified with certainty on this side of the phase diagram. Moreover, we are not aware of any systematic ARPES measurements concerning the nodal Fermi velocity and the Fermi 
momentum as a function of electron-doping to provide the meaning of microscopic parameters and related dynamical processes.

The DE-GWF \cite{Bunemann2012} method used here has been applied to the description of the SC phase in our group quite recently \cite{Kaczmarczyk2013,Kaczmarczyk2014} and subsequently, it has proved to be a useful approach to strongly correlated systems in a number of cases \cite{Wysokinski2015,Wysokinski2016,Kaczmarczyk2016n,zuMunster2016,Zegrodnik2017}. It relies on describing the state of the system by the Gutzwiller-type wave function and allows 
for going beyond the renormalized mean-field theory (RMFT) solution in a systematic manner by using the diagrammatic expansion technique in real space. It has been shown that in practice the first few orders of the expansion suffice to approach asymptotically the full Gutzwiller-wave-function solution with a satisfactory accuracy. Moreover, the method reproduces the VMC results \cite{Kaczmarczyk2014}, it is numerically efficient and is not limited to systems of finite size. Most importantly, as already pointed out, it has led to a very good agreement between experiment and theory for HTS \cite{Spalek2017}.

The paper is organized as follows. In the next Section we present briefly the extended t-J-U model supplemented with both the correlated-hopping and the direct intersite-Coulomb interaction terms, as well as show the basic concepts behind the DE-GWF method as applied to this model. In Section 3 we discuss our results, focusing on the influence of the correlated hopping term on principal features of the  $d$-$wave$ superconducting phase within the approach based on the t-J-U model in the context of copper based HTS. Such a goal should minimally support the claimed universality of our previous results obtained within the t-J-U model, as well as supplement them with new results in the situation when the added terms are important (e.g., when we compare the phase diagram on the hole-doped side with that for the electron-doping). Additionally, at the end of that Section we also discuss selected results coming from the Hubbard model. The conclusions and outlook are the subject of the last Section. In 
Appendix A we supply the most general form of the single-narrow band Hamiltonian for the correlated fermions.

\section{Model and Method}

We start with the t-J-U Hamiltonian \cite{Spalek2017} supplemented with the correlated electron hopping and the direct intersite Coulomb repulsion terms. Its explicit form is (cf. also Appendix)
\begin{equation}
\begin{split}
\mathcal{\hat{H}}&=\sum_{\langle ij\rangle\sigma}(t+K(\hat{n}_{i\bar{\sigma}}+\hat{n}_{j\bar{\sigma}}))\hat{c}^{\dagger}_{i\sigma}\hat{c}_{j\sigma}
+t'\sum_{\langle\langle ij\rangle\rangle\sigma}\hat{c}^{\dagger}_{i\sigma}\hat{c}_{j\sigma}\\
&+J\sum_{\langle ij\rangle}\hat{\mathbf{S}}_i\cdot\hat{\mathbf{S}}_j
+U\sum_i \hat{n}_{i\uparrow}\hat{n}_{i\downarrow}+V\sum_{\langle ij\rangle} \hat{n}_{i}\hat{n}_{j},
 \label{eq:H_start}
 \end{split}
\end{equation}
where the first two terms correspond to the single electron hopping together with the correlated-hopping contribution, the third term represents the antiferromagnetic superexchange interaction, and the last two terms refer to the intra- and inter-site Coulomb repulsion, respectively. By $\langle...\rangle$ and $\langle\langle...\rangle\rangle$ we denote the summations over the nearest-neighbors and next nearest-neighbors, respectively. For $J=0$ one obtains the Hubbard model, which is also briefly discussed later. For $U\rightarrow \infty$ we approach the t-J model limit for which no double occupancies are allowed (cf. Fig. 8 in Ref. \onlinecite{Spalek2017}). In the latter case, the contributions coming from the correlated hopping and from the intrasite Hubbard repulsion vanish for obvious reasons. The physical significance of the t-J-U model was discussed in Ref. \onlinecite{Spalek2017}, together with the determination of the values of microscopic parameters for which the quantitative agreement with 
selected 
experimental data of HTS is achieved. For the sake of clarity and completeness, in Appendix A we discuss the most general form of the narrow-band Hamiltonian, with all two-site terms included. 

It should be noted that due to the presence of the next-nearest neighbor hopping and the correlated hopping terms, the considered Hamiltonian breaks the electron-hole symmetry. In our analysis we use the electron language and define the doping $\delta\equiv 1-n$, where $n$ is the number of electrons per atomic site. Therefore, we obtain $\delta>0$ ($\delta<0$) for the case of hole-doping (electron-doping).

Within the DE-GWF method we assume that the system can be described by the correlated Gutzwiller-type projected many particle wave function of the form
\begin{equation}
 |\Psi_G\rangle\equiv\hat{P}_G|\Psi_0\rangle,
\end{equation}
where $|\Psi_0\rangle$ is the non-correlated wave function subject to our choice. In particular, for the analysis of the SC phase we assume nonzero anomalous averages $\langle\Psi_0|\hat{c}^{\dagger}_{i\uparrow}\hat{c}^{\dagger}_{j\downarrow}|\Psi_0\rangle$, which lead to the $d$-$wave$ pairing amplitude when transformed to reciprocal space. However, due to the the fact that in this approach we include the superconducting averages not only between the nearest neighbors, small corrections to the bare $d$-$wave$ symmetry appear\cite{Kaczmarczyk2014} (see also Sec. III). The general form of the correlation operator $\hat{P}_G$ is provided below 
 \begin{equation}
 \hat{P}_G\equiv\prod_i\hat{P}_i=\prod_i\sum_{\Gamma}\lambda_{i,\Gamma} |\Gamma\rangle_{ii}\langle\Gamma|,
  \label{eq:Gutz_operator}
\end{equation}
where the variational parameters $\lambda_{i,\Gamma}\in\{\lambda_{i\emptyset},\lambda_{i\uparrow},\lambda_{i\downarrow},\lambda_{i d}\}$ correspond to four states of the local basis $|\emptyset\rangle_i\;, |\uparrow\rangle_i\;, |\downarrow\rangle_i\;, |\uparrow\downarrow\rangle_i$ at site $i$, respectively. It has been shown \cite{Bunemann2012} that in order to carry out the diagrammatic expansion efficiently it is convenient to impose the following condition
\begin{equation}
 \hat{P}_i^2\equiv1+x\hat{d}^{\textrm{HF}}_i,
 \label{eq:constraint}
 \end{equation}
 where $x$ is yet another variational parameter and $\hat{d}^{\textrm{HF}}_i\equiv\hat{n}_{i\uparrow}^{\textrm{HF}}\hat{n}_{i\downarrow}^{\textrm{HF}}$, $\hat{n}_{i\sigma}^{\textrm{HF}}\equiv\hat{n}_{i\sigma}-n_{0}$, with $n_{0}\equiv\langle\Psi_0|\hat{n}_{i\sigma}|\Psi_0\rangle$. The $\lambda_{\Gamma}$ parameters are all functions of $x$ which means that there is only one variational parameter of the wave function.
 
 In order to calculate the expectation value of the ground state energy $\langle\mathcal{\hat{H}}\rangle_G\equiv\langle\Psi_G|\mathcal{\hat H}|\Psi_G\rangle/\langle\Psi_G|\Psi_G\rangle$, one can make use of the following relations for any two local operators $\hat{o}_i$ and $\hat{o}^{\prime}_j$ from the Hamiltonian
 \begin{equation}
  \langle\Psi_G|\hat{o}_{i}\hat{o}^{\prime}_{j}|\Psi_G\rangle=\sum_{k=0}^{\infty}\frac{x^k}{k!}\sideset{}{'}\sum_{l_1...l_k}\langle\Psi_0| \tilde{o}_{i}\tilde{o}^{\prime}_{j}\;\hat{d}^{\textrm{HF}}_{l_1...l_k}|\Psi_0 \rangle,
\label{eq:expansion}
\end{equation}
where $\tilde{o}_{i}\equiv\hat{P}_i\hat{o}_{i}\hat{P}_{i}$, $\tilde{o}^{\prime}_{j}\equiv\hat{P}_j\hat{o}^{\prime}_{j}\hat{P}_{j}$, $\hat{d}^{\textrm{HF}}_{l_1...l_k}\equiv\hat{d}^{\textrm{HF}}_{l_1}...\hat{d}^{\textrm{HF}}_{l_k}$, with  $\hat{d}^{\textrm{HF}}_{\varnothing}\equiv 1$. The primmed summation has the restrictions $l_p\neq l_{p'}$, $l_p\neq i,j$ for all $p$ and $p'$. The averages in the non-correlated state on the right-hand side of Eq. (\ref{eq:expansion}) can be decomposed by the use of the Wick's theorem and expressed in terms of the correlation functions $P_{ij} \equiv \langle \hat{c}^{\dagger}_{i\sigma} \hat{c}_{j\sigma}\rangle_0$ and $S_{ij} \equiv \langle \hat{c}^{\dagger}_{i\uparrow} \hat{c}^{\dagger}_{j\downarrow}\rangle_0$. Such a procedure allows us to express the ground state energy $\langle\mathcal{\hat{H}}\rangle_G$ as a function of $P_{ij}$, $S_{ij}$, $n_{0}$, and $x$. It has been shown that the desirable convergence can be achieved by taking the first 
4-6 terms of the expansion in 
$k$ appearing in Eq. (\ref{eq:expansion}). 

In the next step, we derive the effective single-particle Hamiltonian from the minimization condition of the ground-state energy functional $\mathcal{F}\equiv\langle\mathcal{\hat{H}}\rangle_G-\mu_G\langle\hat{N}\rangle_G$, where $\mu_G$ and $\langle N\rangle_G$ are the chemical potential and the total number of particles determined in the state $|\Psi_G\rangle$, respectively. The explicit form of this effective Hamiltonian is
\begin{equation}
 \hat{\mathcal{H}}_{\textrm{eff}}=\sum_{ij\sigma}t^{\textrm{eff}}_{ij}\hat{c}^{\dagger}_{i\sigma}\hat{c}_{j\sigma}+\sum_{ij}\big(\Delta^{\textrm{eff}}_{ij}\hat{c}^{\dagger}_{i\uparrow}\hat{c}^{\dagger}_{j\downarrow}+H.c.\big),
 \label{eq:H_effective}
\end{equation}
where the effective parameters are defined below
\begin{equation}
 t^{\textrm{eff}}_{ij}\equiv \frac{\partial\mathcal{F}}{\partial P_{ij}},\quad \Delta^{\textrm{eff}}_{ij}\equiv \frac{\partial\mathcal{F}}{\partial S_{ij}}.
 \label{eq:effective_param}
\end{equation}
For $i=j$, the $-t^{\textrm{eff}}_{ii}$ has the meaning of the effective chemical potential. It should be noted that when carrying out the calculations, one includes only the non-correlated parameters $P_{ij}$ and $S_{ij}$ which correspond to the relative distances smaller than some arbitrary value, i.e., $|\mathbf{R}_i-\mathbf{R}_j|\leq R_{max}$. In our analysis we have carried our calculations to the fifth order (i.e. $k_{max}=5$) for $R_{max}^2= 10a^2$, with $a$ being the lattice constant. 

Next, the effective Hamiltonian (\ref{eq:H_effective}) is transformed to the reciprocal space and diagonalized, what allows us to derive the self-consistent equations for the quantities $S_{ij}$ and $P_{ij}$. The self-consistent equations are solved numerically together with the concomitant minimization over the $x$ parameter. Additionally, after calculating $P_{ij}$, $S_{ij}$, $x$, $\mu_G$, and $n_0$ for a selected set of microscopic parameters ($t'$, $J$, $U$, $V$), we can determine the value of the so-called correlated SC gaps $\Delta_{G,ij}\equiv\langle \hat{c}^{\dagger}_{i\uparrow} \hat{c}^{\dagger}_{j\downarrow}\rangle_G$. The dominant contribution to the paired phase comes from $\Delta_{G,ij}$ for which $\mathbf{R}_i-\mathbf{R}_j=(1,0)a$ which we denote by $\Delta_G$ for simplicity in the subsequent analysis.

\section{Results and discussion}
In our analysis we have selected the values of $t$, $t'$, $J$, $U$ close to those in Ref. \onlinecite{Spalek2017} as they lead to a very good agreement with the selected experimental data. Namely, we set $t=-0.35$ eV, $t'=0.3|t|$, $J=0.25|t|$, $U=21|t|$. The intersite Coulomb repulsion term is varied within the interval $V\in [\;0,\;2|t|\;]$. According to the estimates of $K$ for the $3d$ electrons, the ratio $K/U$ should be taken as relatively small \cite{Hubbard1963}. In our analysis we limit to the values between $0$ and $0.5|t|$. In the subsequent discussion all the energies are defined in units of $|t|$, unless stated otherwise. The calculations have been carried out for the case of hole doping ($\delta>0$, $\delta\equiv 1-n$) and the electron doping ($\delta<0$). The so-called BCS-like and non-BCS superconducting regimes which are discussed in this Section are defined by the sign of the kinetic energy gain at the transition to the SC phase, $\Delta E_{\mathrm{kin}}$. Namely, for the BCS-like regime we 
have $\Delta E_{\mathrm{
kin}}
>0$ and 
for the non-BCS state $\Delta E_{\mathrm{kin}}<0$. The kinetic energy gain is defined explicitly in the following manner
\begin{equation}
\Delta E_{\mathrm{kin}}=E^{SC}_{G|0}-E^{PM}_{G|0},\quad E_{G|0}\equiv \frac{1}{N}\sum_{ij\sigma}t_{ij}\langle\hat{c}^{\dagger}_{i\sigma}\hat{c}_{j\sigma}\rangle_G,
\end{equation}
where SC and PM superscripts correspond to the superconducting and paramagnetic phases, respectively, and the averages are of course calculated for the Gutzwiller state, $|\Psi_G\rangle$.

\begin{figure}[h!]
\centering
\epsfxsize=80mm 
\epsfbox[115 500 400 773]{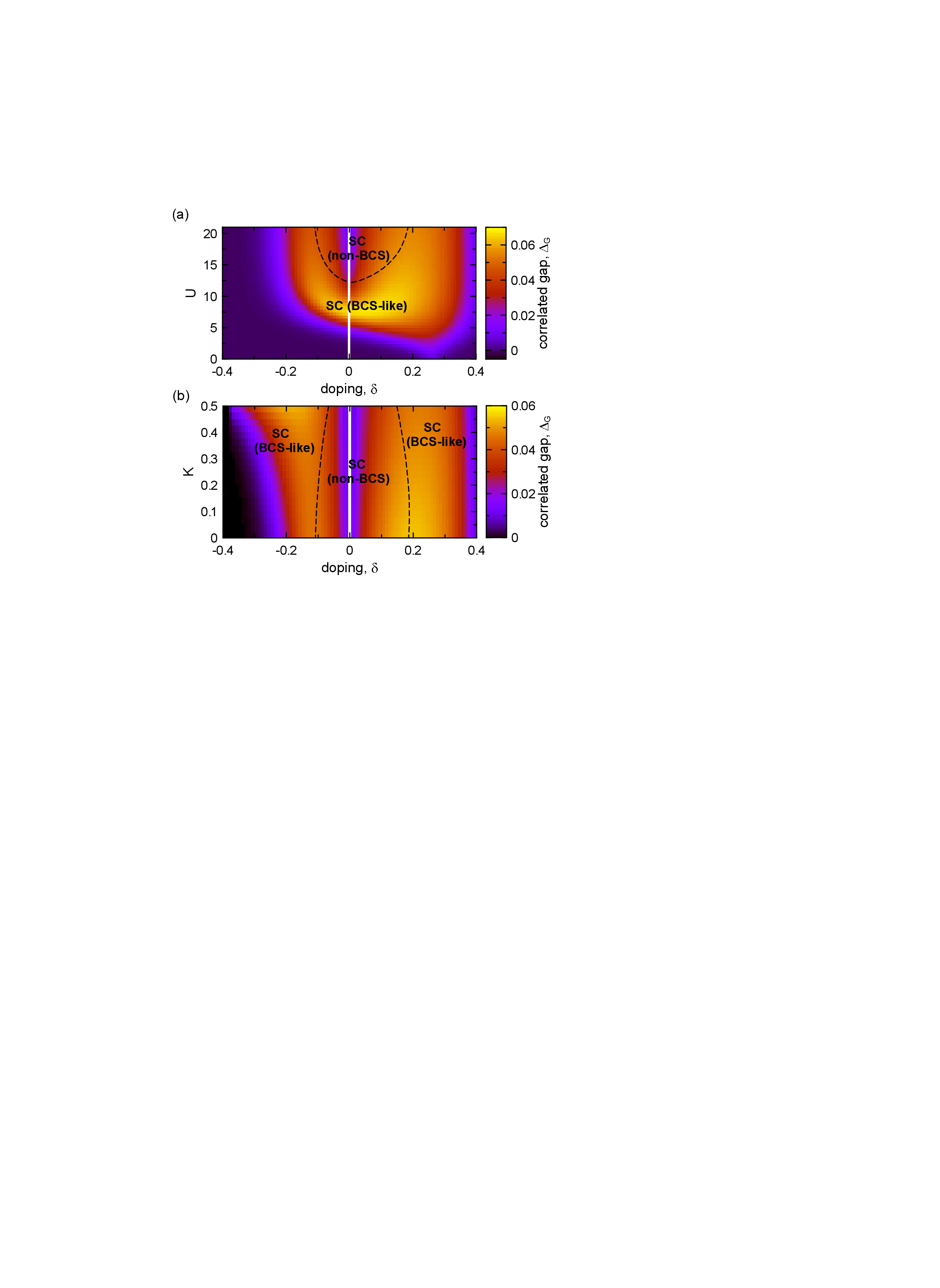}
\caption{(Colors online) (a) Correlated gap ($\Delta_G$) as a function of intrasite Coulomb repulsion $U$ and doping $\delta$, for $J=0.25$, $V=0$. (b) Correlated gap as a function of correlated hopping integral $K$ and $\delta$, for $J=0.25$ and $U=21$, $V=0$. In both figures the borders between the BCS-like and non-BCS regimes are marked by the dashed lines.}
\label{fig:diags_tJU}
\end{figure}
In Fig. \ref{fig:diags_tJU}a we show the location of the non-BCS superconducting regime on the doping $\delta$-interaction $U$ plane for the case of the t-J-U model. The correlated gap values for each $(\delta, U)$ point on the phase diagram are provided by the colored scale. It can be seen that for high enough values of the Coulomb repulsion the non-BCS phase is contained within the underdoped regime $\delta<\delta^h_{\mathrm{opt}}$ ($\delta>\delta^e_{\mathrm{opt}}$) for the hole (electron) doping. The values of the optimal dopings $\delta^h_{\mathrm{opt}}$ and $\delta^e_{\mathrm{opt}}$ are evaluated as the dopings for which the maximum of the correlated gap ($\Delta_G$) appears in the hole and electron doped situations, respectively. The appearance of the non-BCS state in the underdoped regime obtained by us is in agreement with the experimental reports \cite{Deutscher2005,Carbonne2006,Molegraaf2002,Gianetti2011} for the case of the hole-doping. As we have shown in Ref. \onlinecite{Spalek2017}, the 
incorporation of the $J$ term 
into the t-J-U model, as well as the inclusion of the higher order terms in the DE-GWF method, is of crucial importance in obtaining the proper agreement between the theoretical results and experimental data concerning the non-BCS regime appearance in the hole-doped case. Unfortunately, to the best of our knowledge there are no corresponding experimental data which would show the doping dependence of the kinetic energy gain on the electron-doping (left) side of the phase diagram. 

In Fig. 1b we show the influence of the correlated electron hopping term for $U=21$. As one can see, the value of doping at which the crossover between the BCS-like and non-BCS regimes appears moves slightly towards the half-filling ($\delta=0$) with the increasing $K$. On the other hand, $\delta^{e}_{\mathrm{opt}}$ is moving away from the half-filling with a slight increase of the correlated gap values while the $K$ parameter is increased. A similar behavior has been reported for the case of the Hubbard model, cf. Ref. \onlinecite{
Wysokinski2017}. The 
effect 
of the correlated hopping term for $\delta<0$ leads to the situation in which the crossover between the two regimes is no longer located in the vicinity of the the optimal doping. However, on the hole-doping side the effect of $K$ is not that significant and $\delta^{h}_{\mathrm{opt}}$ is still close to $\delta_c$ for which the crossover appears even for a relatively high values of $K$. Note also that only for relatively small $K\lesssim 0.2$ the upper critical concentration for the disappearance of HTS state on the electron side is significantly lower than that for the hole doped case, in agreement with experiment \cite{Keimer2015}.

\begin{figure}[h!]
\centering
\epsfxsize=80mm 
\epsfbox[236 341 497 729]{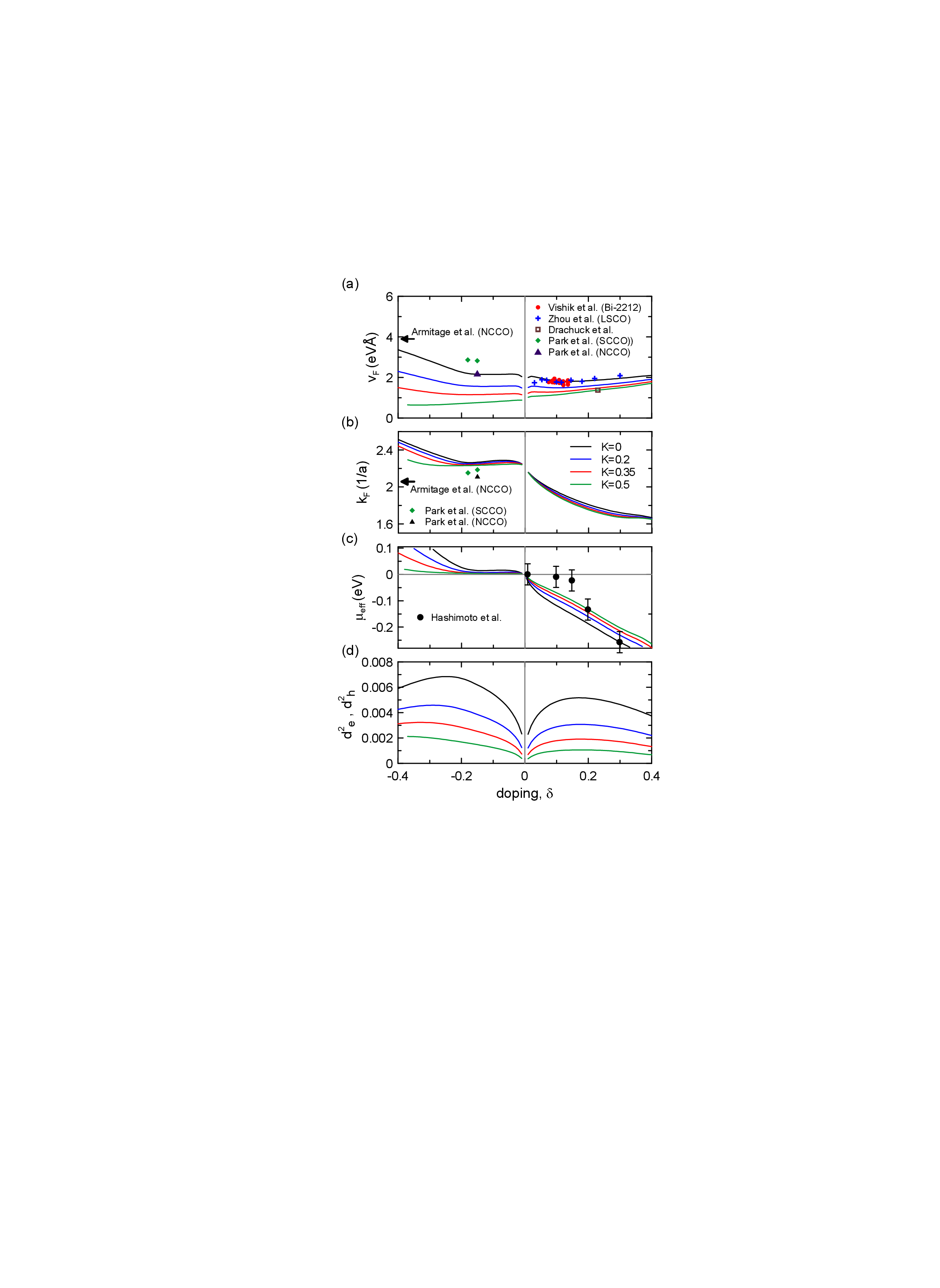}
\caption{(Colors online) Doping dependences of Fermi velocity in the nodal direction (a), nodal Fermi momentum (b), effective chemical potential (c), and double occupancy of electrones $d^2_e$ (at the hole-doping side) and holes $d^2_h$ (at the electron-doping side) (d), for selected values of the correlated-hopping integral $K$. In (a) we show also the measured values of $v_F$ which are taken from Refs. \onlinecite{Vishik2010,Zhou2003,Drachuck2014,Park2008,Armitage2003}. For the case of electron-doping the values of $v_F$ have been extracted by us from the measured dispersion relations close to $E_F$ presented in Ref. \onlinecite{Park2008,Armitage2003}. In Ref. \onlinecite{Armitage2003} the doping of the sample has not been explicitly provided so the corresponding $v_F$ value is marked by us by an arrow on the vertical scale. In analogy to (a), we show in (b) also the Fermi momentum values at the electron doped side extracted from Refs. \onlinecite{Park2008,Armitage2003} In (c) we show the measured values 
of the chemical potential shift for LSCO compound taken from Ref. \onlinecite{Hashimoto2008}. The calculations have been carried out for $J=0.25$, 
$U=21$, and $V=0$.}
\label{fig:v_k_d}
\end{figure}

\begin{figure}[h!]
\centering
\epsfxsize=89mm 
\epsfbox[250 333 550 693]{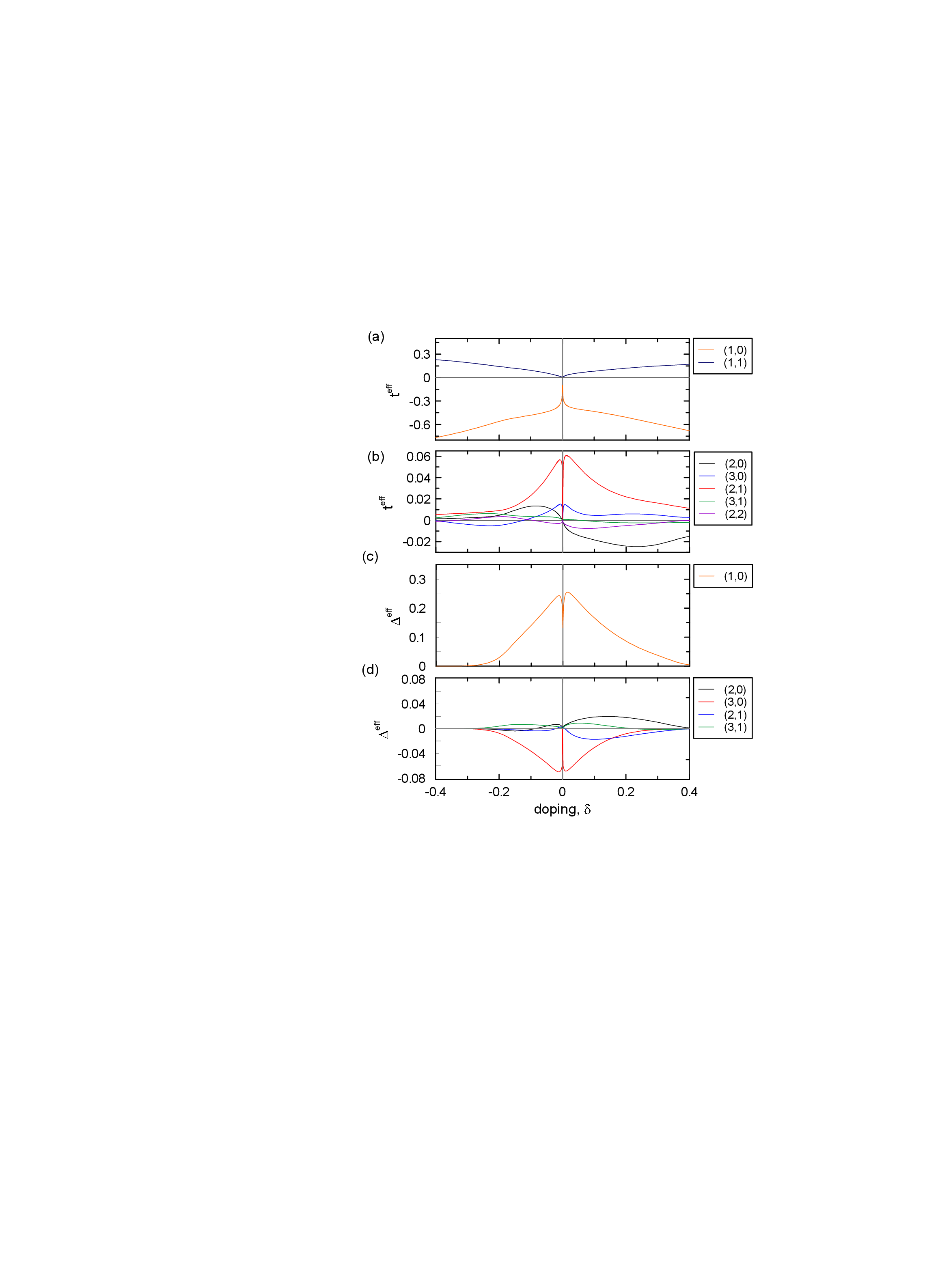}
\caption{(Colors online) (a), (b) Doping dependence of the effective hopping parameters $t^{\mathrm{eff}}_{ij}$ between subsequent neighboring sites defined by the vectors $\Delta \mathbf{R}_{ij}=\mathbf{R}_i-\mathbf{R}_j=(X,Y)$. The vector components provided in the Figure are given in the units of lattice constant $a$. (c), (d) The corresponding effective gap magnitudes $\Delta^{\mathrm{eff}}$. The results correspond to $J=0.25$, $U=21$, and $K=0$.}
\label{fig:add}
\end{figure}

In Fig. \ref{fig:v_k_d} we show the nodal Fermi velocity (a), nodal Fermi momentum (b), effective chemical potential (c), as well as double occupancy (d) changes caused by the correlated hopping term. We see that the nodal Fermi velocity is decreased with increasing $K$. The APRES measurements for the hole-doped cuprates in the range 30-40 meV \cite{Vishik2010} give the value of $v_F\approx 2.0$ eV \AA$\;$ which is practically independent of the doping. Similar doping independence has been reported in Ref. \onlinecite{Zhou2003}. These data are in good agreement with our theoretical results for small values of $K$ (cf. also Ref. \onlinecite{Spalek2017}). However, one should note that an essentially lower value of the Fermi velocity $v_F\approx (1.36\pm 0.06)$ eV\AA$\;$ has been observed in more complicated cuprate superconductor \cite{Drachuck2014}. Also, the nodal Fermi velocity measured very close to the Fermi energy (within 7 meV) is observed to decrease remarkably as we are approaching the half filling 
from the hole-doped side of the phase diagram \cite{Vishik2010}. This effect is caused by a low-energy kink in the dispersion relation which is of unknown origin. Such kink is not reproduced within our approach. Therefore, the latter behavior cannot be interpreted in the context of the presented results. As we have already mentioned the detailed analysis of the experimental electron-doping dependence of the Fermi velocity has not been determined convincingly as yet. However, by extracting the Fermi velocity from the available ARPES data\cite{Park2008,Armitage2003} for the NCCO and SCCO compounds we obtain the values $v_F\approx 2$ eV\AA$\;$ and $v_F\approx 3$ eV\AA, respectively for $\delta=-0.15$ (and $\delta=-0.18$ for SCCO). However, data provided in Ref. \onlinecite{Armitage2003} lead to larger values of $v_F\approx 4$ eV\AA$\;$ for NCCO. 
Our calculations for the electron doped HTS lead to small doping dependence of the Fermi velocity for $\delta\gtrsim -0.2$ with $v_F\approx 2$ eV\AA$\;$ which corresponds to the experimental value for NCCO from Ref. \onlinecite{Park2008}.

From Fig. \ref{fig:v_k_d} one can see that the Fermi momentum $k_F$ is not drastically affected by $K$. For $K=0$ and for the case of electron-doping, $k_F$ is very weakly dependent on $\delta$. The calculated values of $k_F$ roughly agree with those seen in experiment for NCCO and SCCO\cite{Park2008,Armitage2003}. On the hole-doped side small, but clearly visible increase of $k_F$ appears when $\delta$ approaches the half-filling. Similar increase of $k_F$ is also seen in experiment for LSCO as shown in Ref. \onlinecite{Hashimoto2008} (cf. Ref. \onlinecite{Spalek2017}). From Figs. \ref{fig:v_k_d} b and c one can see that the Fermi momentum reflects the behavior of the effective chemical potential [$\mu^{\mathrm{eff}}=-\partial\mathcal{F}/\partial P_{ii}\;$, cf. (\ref{eq:H_effective})]. In Fig. \ref{fig:v_k_d} c we have provided also the measured values of the chemical potential shift for LSCO high temperature superconductor, which are taken from Ref. \onlinecite{Hashimoto2008}. 

The electron ($d_e^2\equiv \langle\hat{n}_{i\uparrow}\hat{n}_{i\downarrow}\rangle_G$) and hole ($d^2_h=d^2_e-\delta$) double occupancies shown in Fig. \ref{fig:v_k_d}d are both suppressed by the correlated hopping term. This stems from the fact that the more double occupancies are in the system, the more correlated hopping events may take place [cf. Eq. (\ref{eq:H_start})], which in turn increases the system energy (since $K$ is positive). As a result, the number 
of double occupancies is suppressed to reduce the system energy. In the limiting situation of $d_e^2\equiv 0$, which corresponds to the t-J model, the contribution to the system energy resulting from the correlated hopping term is exactly zero. 

It should be noted that even though in the starting Hamiltonian we take only the nearest ($t=-0.35$ eV) and next-nearest neighbor hoppings ($t'=0.3|t|$) as nonzero, in the effective Hamiltonian the hopping parameters corresponding to more distant sites also appear. This comes as a result of the correlation effects of increased range taken in the higher orders of the diagrammatic expansion. However, the nearest- and next-nearest neighboring effective hopping parameters, $t^{\mathrm{eff}}_{10}$ and $t^{\mathrm{eff}}_{11}$ (shown in Fig. \ref{fig:add} a), are still dominant and have one order of magnitude larger values than the remaining ones (shown in Fig. \ref{fig:add} b). As we approach the half-filled situation all the effective parameters tend to zero what illustrates the Mott insulating state being approached in that limit. 

In Fig. \ref{fig:add} we provide the effective gap components, both the dominant (c) and the minor (d). We see that the short-range interaction induces the spin-singlet correlations on a longer scale, in accordance with the idea of resonating valence bond (RVB) proposed originally for the spin-liquid state by Anderson and Fazekas\cite{Anderson1973,Fazekas1974,Randeria2016}. One may thus see that the present approach extends the RVB concept to the spin-singlet superconducting state. It must be noted that the effective gap magnitude $\Delta$ in the antinodal direction, as it comes from the dispersion relation in SC phase, reproduces only quantitatively the respective data trend, as has been discussed earlier within the Hubbard\cite{Kaczmarczyk2013} and t-J models\cite{Kaczmarczyk2014}. The question of difference between the antinodal gap and the gap determined from the slope of the dispersion relation close to the nodal point\cite{Kaminski2007,Yoshida2012,Vishik2012} cannot be resolved within this approach.  
One important feature of the results in Fig. \ref{fig:add} c and d is that we have plotted only some of the gap components. Namely, the plotted $(n,m)$ component has also its $(m,n)$ counterpart of the same amplitude, but of the opposite sign. In effect, the gap has the form of the $d-wave$, with an admixture of smaller-amplitude higher Fourier harmonics.

\begin{figure}[h!]
\centering
\epsfxsize=85mm 
\epsfbox[196 583 485 782]{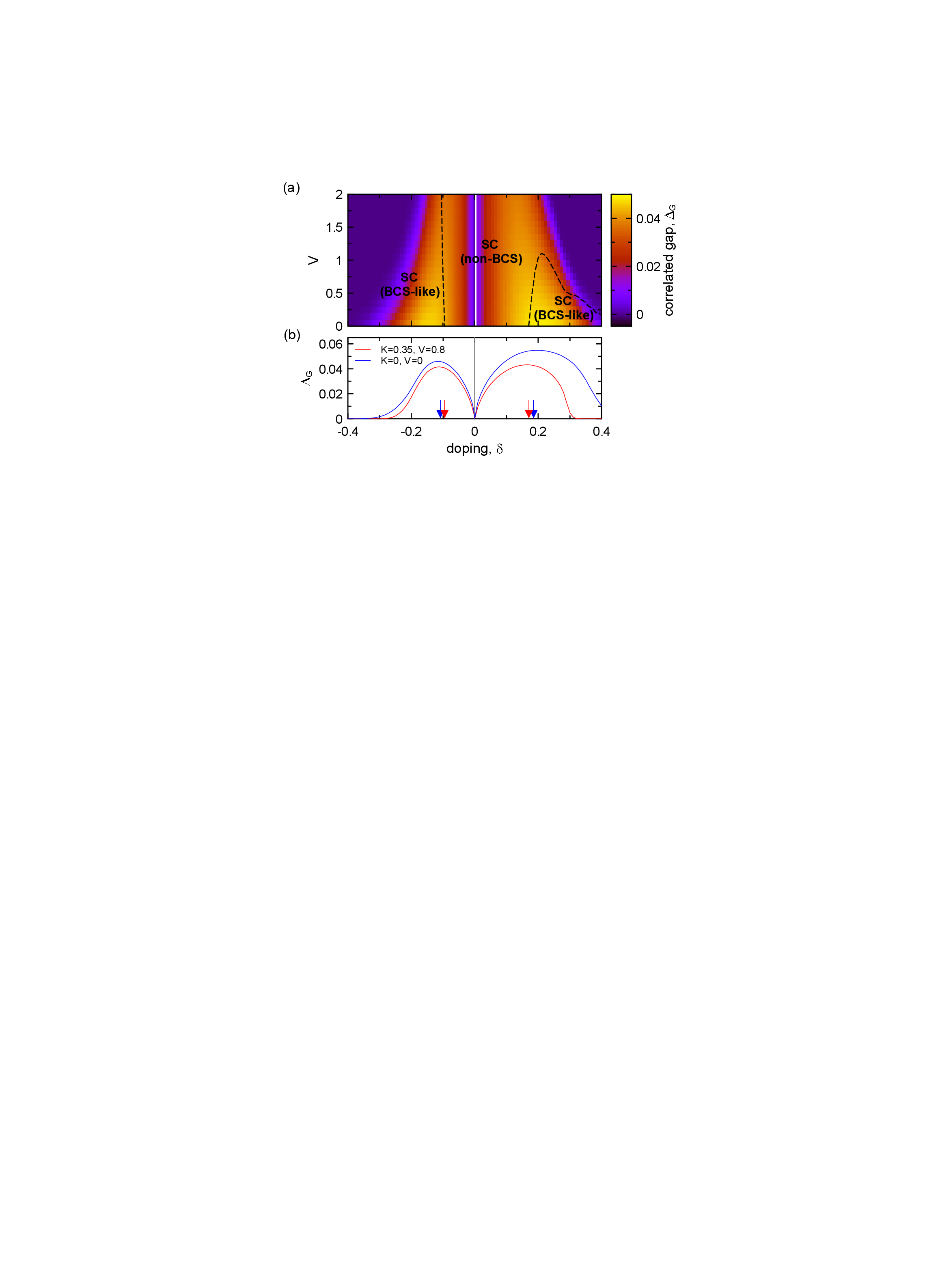}
\caption{(Colors online) (a) Correlated gap as a function of intersite Coulomb repulsion $V$ and doping $\delta$, for $J=0.25$, $U=21$, and $K=0.35$. The borders between the BCS-like and non-BCS regimes are marked by the dashed lines. (b) Correlated gap as a function of doping for two selected sets of model parameters. The vertical arrows show the dopings for which the crossover between the BCS-like and non-BCS states appear for the corresponding model parameters.}
\label{fig:diag_nV}
\end{figure}

As the correlated hopping term comes as a result of the off-diagonal element of
the Coulomb interaction between the nearest neighboring lattice sites, we supplement also our model with the intersite Coulomb repulsion $\sim V$ [the last term in Hamiltonian (\ref{eq:H_start})]. The results for the case of $K=0.35$ and increasing $V$ are provided in Fig. \ref{fig:diag_nV} a. As one can see, the influence of intersite Coulomb repulsion for the case of nonzero $K$ and for the hole-doping case is similar as for $K=0$ which has been discussed in Ref. \onlinecite{Abram2016}. Namely, the upper critical doping for the SC-phase disappearance ($\delta_c$) is moving towards the half-filling and the non-BCS regime expands with increasing $V$. Nevertheless, for $V\lesssim 1$ the crossover between the BCS-like and non-BCS phases is still quite close to the optimal doping. For the electron-doped case the rise of $V$ also results in moving $\delta_c$ towards half-filling. However, the doping value which corresponds to the crossover between BCS-like and non-BCS states is practically not affected. 

\begin{figure}[h!]
\centering
\epsfxsize=85mm 
\epsfbox[176 350 478 737]{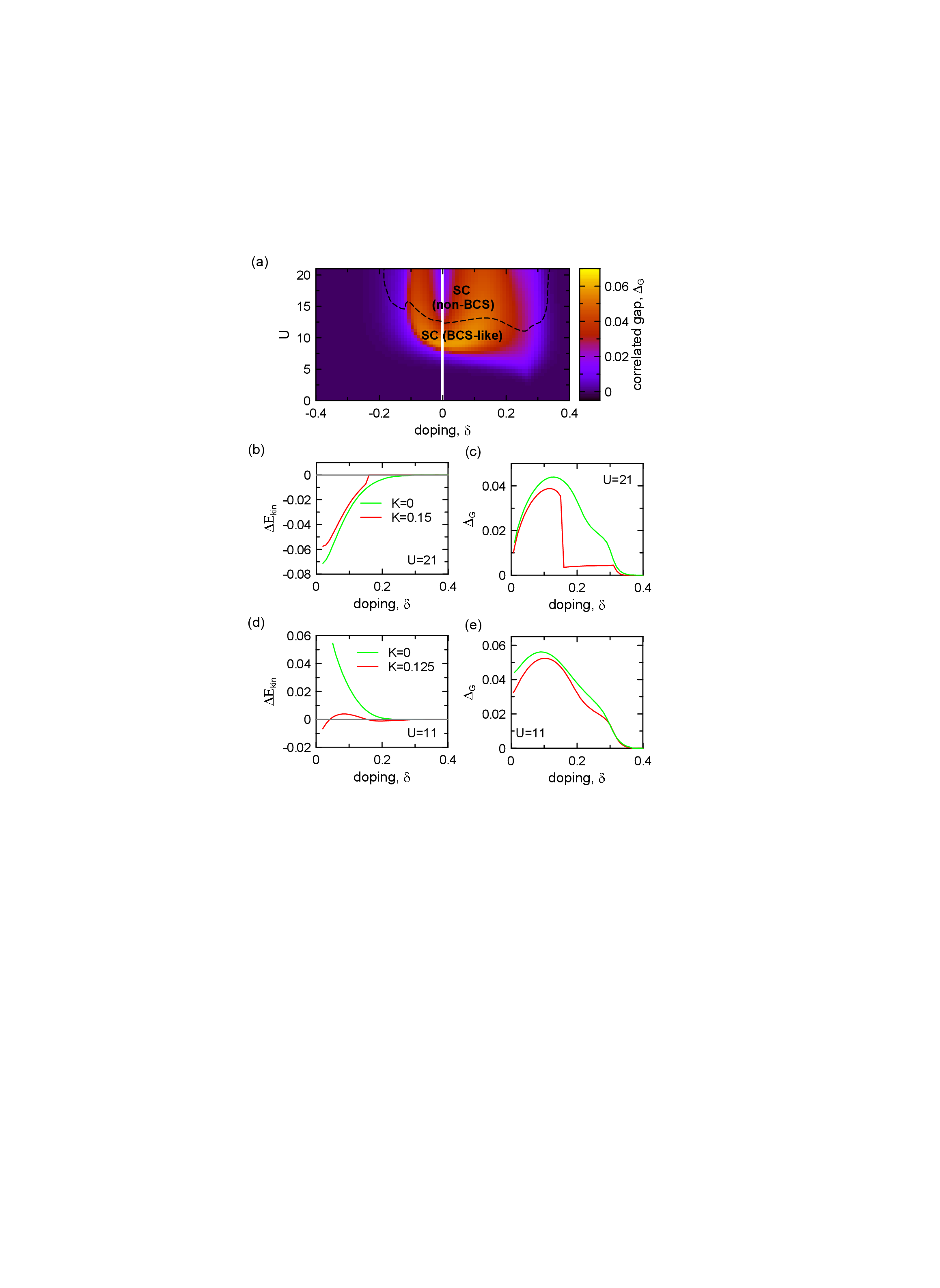}
\caption{(Colors online) (a) Correlated gap as a function of both $U$ and $\delta$ for the Hubbard model. The border between the BCS-like and non-BCS regimes is marked by the dashed line. In (b) and (c) we show the kinetic energy gain at the transition to the superconducting phase, as well as the correlated gap as a function of doping for $U=21$ and for the two selected values of $K$ ($K=0.0$, $K=0.15$). In (d) and (e) we show also $\Delta E_{\mathrm{kin}}$ and $\Delta_G$ as a function of doping for a significantly lower value of $U$ ($U=11$) and for the two selected values of $K$ ($K=0.0$, $K=0.125$). }
\label{fig:diag_Hub}
\end{figure}

For the sake of completeness, in Fig. \ref{fig:diag_Hub} we display also the selected results corresponding to the Hubbard model [$J=0$ and $V=0$ in (\ref{eq:H_start})]. In Fig. \ref{fig:diag_Hub}a we show the phase diagram on the $(\delta,U)$ plane which shows that above the value of $U\approx 12$ non-BCS regime appears in the wide range of dopings and the crossover between the non-BCS and BCS-like phases does not appear anywhere close to the optimal doping, in contradiction to the experimental data \cite{Deutscher2005,Carbonne2006,Molegraaf2002,Gianetti2011} and the results obtained within the t-J-U model (cf. Fig. \ref{fig:diags_tJU}a here and Ref. \onlinecite{Spalek2017}). Moreover, below the value $U\approx 12$ the BCS-like regime appears within the whole doping range of SC phase stability. This is also in disagreement with experiment. In Fig. \ref{fig:diag_Hub}b we show that with taking nonzero value of $K$, the sign of $\Delta E_{\mathrm{kin}}$ does not change and 
the non-BCS state still extends over a wide doping range in the high-U regime. However, a sudden drop in the correlated gap as a function of doping is induced by the correlated hopping as shown in Fig. \ref{fig:diag_Hub}c. Such a behavior signals the appearance of phase separation and it has also been reported in Ref. \onlinecite{Wysokinski2017} for the electron-doped system but for significantly smaller values of $U$. On the other hand, the t-J-U model does not show such feature (both for electron- and hole-doping cases) even for significantly larger values of $K$ (cf. Fig. \ref{fig:diags_tJU}b).

In Fig. \ref{fig:diag_Hub}d we show that the sign change of the kinetic energy gain can be induced by the correlated hopping below $U\approx 12$, where originally only the BCS-like phase appears with positive $\Delta E_{\mathrm{kin}}$ (cf. Fig. \ref{fig:diag_Hub}a). As shown in Fig. \ref{fig:diag_Hub}e, the correlated gap is slightly reduced by the 
nonzero value of $K$, however, no phase separation appears.

\begin{figure}[h!]
\centering
\epsfxsize=75mm 
\epsfbox[290 624 531 760]{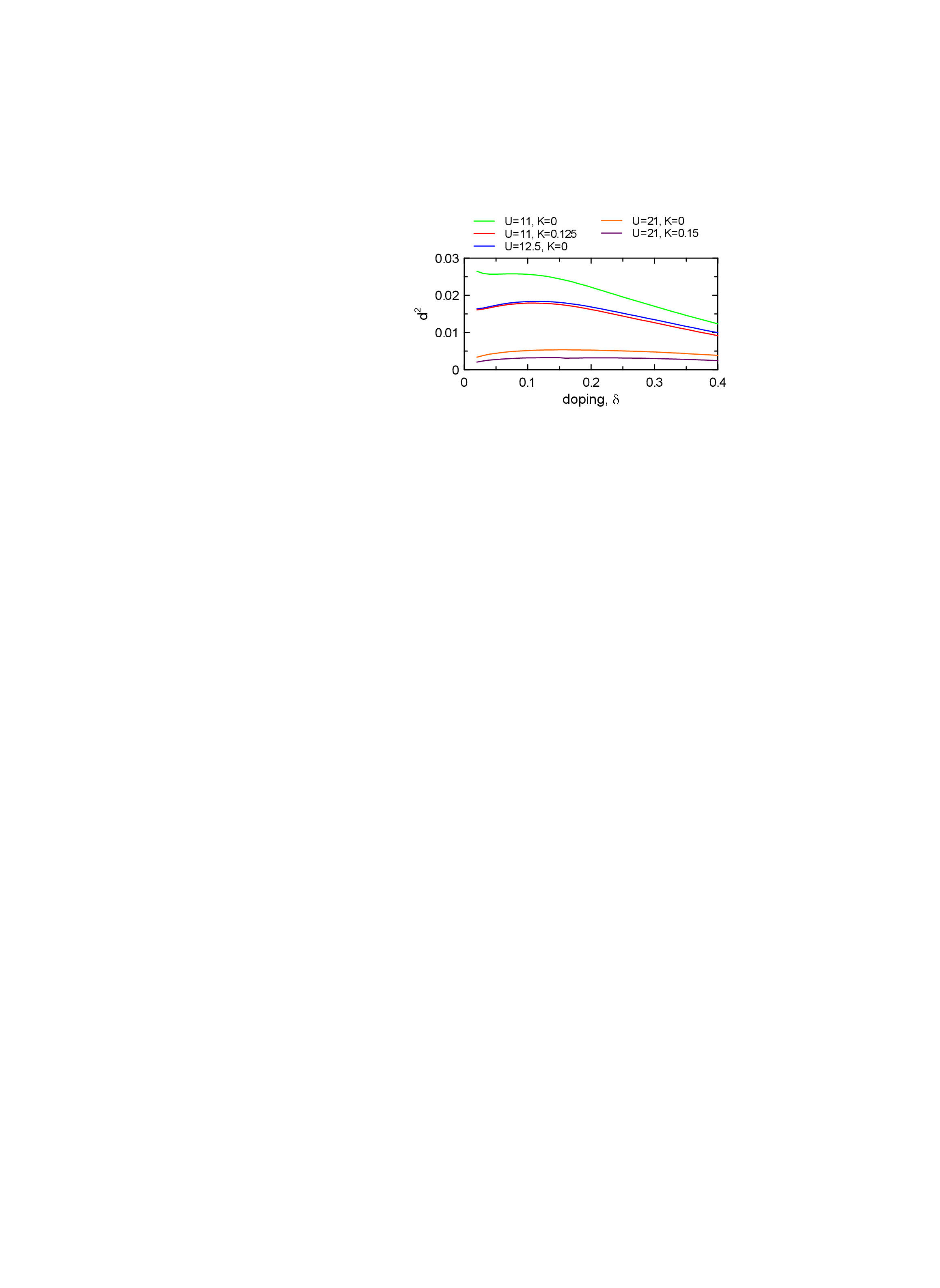}
\caption{(Colors online) Double occupancy as a function of doping for selected interaction parameters $U$ and $K$ in the case of the Hubbard model ($J=0$, $V=0$).}
\label{fig:double_occupancy_Hubb}
\end{figure}

Finally, in Fig. \ref{fig:double_occupancy_Hubb} we have plotted double occupancy of electrons as a function of doping for selected values of $U$ and $K$. We can see that the average number of double occupancies in the system for a given value of $U$ ($U=12.5$) and $K=0$ can be very close to those corresponding to smaller value of $U$ and nonzero $K$ (e.g., $U=11$ and $K=0.125$). For $U=21$ one obtains the number of double occupancies one order of magnitude smaller than that for $U\approx 11$.

\section{Conclusions and outlook}

We have analyzed the effect of the correlated hopping term $\sim K$ and intersite Coulomb repulsion $\sim V$ on the $d$-$wave$ superconductivity within the t-J-U model by using the DE-GWF method in both the electron- and the hole-doping regimes. We show that by increasing $K$ we decrease the doping range in which the so-called non-BCS phase appears (see Fig. \ref{fig:diags_tJU}). At the same time, the optimal doping moves away from the half filling. Nevertheless, for the case of hole doping the crossover between the BCS-like and non-BCS regimes is still close to the optimal doping, even for reasonably large value of $K$, as it is seen in experiment \cite{Deutscher2005,Carbonne2006,Molegraaf2002,Gianetti2011}. This is not the case for $\delta<0$ (the electron doping), where the change in the optimal doping is much more pronounced by the presence of the $K$ term. However, the experimental data concerning the location of the non-BCS regime for the electron-doping situation have not been presented so far, to the 
best of our knowledge. It 
would be important to test 
the validity of this difference, as it would show (if any) the influence of the electron-hole asymmetry.

The calculated nodal Fermi velocity for low values of $K$ show a very weak doping dependence for $-0.2\lesssim \delta\lesssim 0.4$, in agreement with the available experimental data for the hole-doped high temperature superconductors\cite{Zhou2003,Vishik2010}. The selected experimental value of $v_F\approx 2$ eV\AA$\;$ on the electron-doped side for NCCO taken from Ref. \onlinecite{Park2008} also agrees well with our results. However, reports of larger values of $v_F\approx 3-4$ eV\AA$\;$ for electron dopings are also available \cite{Park2008,Armitage2003} (see Fig. \ref{fig:v_k_d}). According to our calculations such higher $v_F$ values appear only below $\delta\approx -0.2$. Furthermore, as the strength of the correlated hopping term is increased, the nodal Fermi velocity is suppressed leading to the decrease in $v_F$ with the decreasing doping (see Fig. \ref{fig:v_k_d}a). 

According to our analysis the nodal Fermi momentum is not significantly affected by the correlated hopping in the whole doping range considered (see Fig. \ref{fig:v_k_d}b). The obtained doping dependence of $k_F$ approximately agrees with the experiment (for LSCO\cite{Hashimoto2008,Spalek2017} as well as SCCO and NCCO\cite{Park2008}). Our results show the pinning of the chemical potential at the electron-doped side of the diagram, where in spite of changes in $\delta$, the value of $\mu_{\mathrm{eff}}$ is practically unchanged, especially for $\delta>-0.2$. For the case of hole-doping, the values of $\mu_{\mathrm{eff}}$ are roughly in agreement with the measured chemical potential shift provided in Ref. \onlinecite{Hashimoto2008}. However, the measured values drop almost to zero for $\delta\lesssim 0.2$ what is not reproduced within our approach where approximately linear behavior of $\mu_{\mathrm{eff}}$ is obtained. From obvious reasons the correlated hopping decreases the double occupancy in the system \
cite{Marsiglio}.

The $V$-term suppresses the paired phase both when $K=0$ (shown in Ref. \onlinecite{Abram2016}) and $K\neq 0$ (analyzed here). For values of $V\approx 1-2$ the upper critical doping for the disappearance of the SC is significantly reduced (Fig. \ref{fig:diag_nV}), in accordance with the experiment. The influence on the non-BCS phase is unusual as for the case of hole doping the non-BCS phase is enhanced with the increasing $K$ and for the electron-doping no significant changes in this respect are reported (cf. Fig. \ref{fig:diag_nV}).

It has been emphasized before \cite{Spalek2017}, and also presented here (cf. Fig. \ref{fig:diag_Hub}a), that in the absence of the correlated hopping term the Hubbard model gives the BCS-like behavior in the wide doping range for $U\lesssim 12$ and the non-BCS type of paired state for high $U$ values ($U\gtrsim 12$). However, as we have shown here the non-BCS behavior can be obtained also below $U\approx 12$ after inclusion of the correlated-hopping term (cf. Fig. \ref{fig:diag_Hub}d). Such a situation can be understood in the following manner. The Hubbard $U$ determines the number of double occupancies in the system. From the phase diagram displayed in Fig. \ref{fig:diag_Hub} one can see that the number of double occupancies have to be small enough to make it possible for the non-BCS phase to appear. The $K$ term additionally suppresses $d^2$ leading to the non-BCS behavior even below $U\approx 12$. It is shown in Fig. \ref{fig:double_occupancy_Hubb} that the double occupancy values for ($U=12.
5$, $K=0$) 
and ($U=11$, $K=0.125$) are very close in the whole doping range. For both these sets of parameters the non-BCS phase appears (cf. Figs. \ref{fig:diag_Hub} a and b), whereas for the case of ($U=11$, $K=0$) the double occupancies are visibly larger which leads only to the BCS-like behavior (Fig. \ref{fig:diag_Hub} a and b). 
However, one should note that the number of double occupancies is not the only parameter which is important in the context of the non-BCS state appearance. Namely, the exchange interaction term, absent in the Hubbard model, plays an important role in obtaining the range of the non-BCS behavior which is consistent with experiment, as shown by us in Ref. \onlinecite{Spalek2017}.

It should be emphasized that the electron-hole symmetry breaking induced by the correlated hopping term in the Hubbard model has also been analyzed recently in Ref. \onlinecite{Wysokinski2017}. However, the effect of sign change of the kinetic energy gain (leading to the appearance of the non-BCS phase) upon the increase of $K$ has not been reported there for the case with $t'\neq 0$. 

In general, relatively small values of $V$ and $K$ ($<1$) introduced here must be assumed to uphold our quantitative picture obtained in Refs. \onlinecite{Spalek2017,Zegrodnik2017}. On the other hand, some systems show that the $v_F$ value is decisively smaller than 2 eV\AA \cite{Drachuck2014}. Therefore, it is conceivable that a delicate interplay between the microscopic parameters takes place to obtain the agreement between experiment and theory for a number of copper-based compounds. The situation for the electron-doped systems is not established as yet to the degree allowing for a systematic quantitative comparison between the two.

\section{Acknowledgement}

We acknowledge the financial support through the Grant MAESTRO, No. DEC-2012/04/A/ST3/00342 from the National Science Centre (NCN) of Poland. The authors are also grateful to Prof. Amit Keren from Technion (Haifa) for insightful remarks and discussion on universality of the Fermi velocity.
\newline
\newline

\appendix
\section{General form of the Hamiltonian of electrons in a narrow band with all two-site interactions}
For the sake of comparison with our starting model defined by Eq. (\ref{eq:H_start}) we supplement the paper with the most general Hamiltonian for interacting electrons in a single narrow band with all the two-site interaction. We start with the proper second-quantization representation, namely
\begin{multline}
\mathcal{\hat H} = \sum_{\sigma} \int d^3\!{\bf r}\, \widehat{\Psi}_{\sigma}^{\dagger}({\bf r}) H_1({\bf r})\, \widehat{\Psi}_{\sigma}({\bf r}) + \\
+ \frac{1}{2} \sum_{\sigma_{1}\sigma_{2}} \iint d^3\! {\bf r}_1\,d^3\!{\bf r}_2\,\widehat{\Psi}_{\sigma_1}^{\dagger}({\bf r}_{1})\, \widehat{\Psi}_{\sigma_2}^{\dagger}({\bf r}_{2})\, V\!\left({\bf r}_{1}-{\bf r}_{2}\right)\\
\times
\widehat{\Psi}_{\sigma_2}({\bf r}_2)\widehat{\Psi}_{\sigma_1} ({\bf r}_1).
\label{eq:general_Ham}
\end{multline}
This form does not include any explicit spin-dependent term in either single-particle or two particle terms taken from the wave mechanics. The field operator, $\widehat{\Psi}_{\sigma}(\mathbf{r})$, for particles with spin $\sigma=\pm 1$ is taken in the following form  
\begin{equation}
 \widehat{\Psi}_{\sigma}(\mathbf{r})=\sum_{i}\phi_i(\mathbf{r})\chi_{\sigma}\hat{c}_{i\sigma},
 \label{eq:field_operator}
\end{equation}
where $\phi_i(\mathbf{r})\equiv\phi(\mathbf{r}-\mathbf{R}_i)$ is the Wannier function centered on site $\mathbf{R}_i$, $\chi_{\sigma}$ is its spin part for the case with the global spin quantization axis, and $\hat{a}_{i\sigma}$ is the annihilation operator of fermion in the single-particle state $|i\sigma \rangle\equiv|\phi_i(\mathbf{r})\chi_{\sigma}\rangle$.

Substituting (\ref{eq:field_operator}) and its Hermitian conjugate counterpart $\Psi^{\dagger}_{\sigma}(\mathbf{r})$ in (\ref{eq:general_Ham}) we obtain
\begin{equation}
\mathcal{\hat H} =\sum_{ij\sigma } t_{ij}\, \hat c_{i\sigma}^{\dagger}\, \hat c_{j\sigma }
+ \, \frac{1}{2} \sum_{ijkl} V_{ijkl} \sum_{\sigma_{1}\sigma_{2}} \hat c_{i\sigma_1}^{\dagger}\hat c_{j\sigma_2}^{\dagger} \hat c_{l\sigma_2}\, \hat c_{k\sigma_1},
\label{eq:2q_Ham}
\end{equation}
where
\begin{equation}
t_{ij} \equiv \int d^3\!{\bf r}\,\Phi_{i}^{*}({\bf r})\, H_{1} ({\bf r})\,\Phi_{j}({\bf r}),
\label{eq:tij}
\end{equation}
\begin{equation}
 V_{ijkl} \equiv \int \Psi_i^*({\bf r}_1) \Psi_j^*({\bf r}_2) V({\bf r}_1 - {\bf r}_2) \Psi_k({\bf r}_1) \Psi_l({\bf r}_2).
 \label{Vijkl}
\end{equation}
The terms appearing in (\ref{eq:2q_Ham}) can be classified as single-site ($i=j=k=l$) , two-site ($i=j\neq k=l$, $i=k\neq j=l$, $i=l\neq j=k$, and $i=j=k\neq l$, etc.), and the remaining three- and four-site contributions. The last two types of terms are usually neglected, although they can play a significant role in the low-dimensional systems, i.e., when the screening effects are ineffective (e.g. for systems of nanoscopic size \cite{Biborski2016}). The single site terms are the following
\begin{equation}
 I_1=\sum_{i\sigma}t_{ii}\hat{n}_{i} + \frac{1}{2}\sum_{i\sigma}V_{iiii}\hat{n}_{i\sigma}\hat{n}_{i\bar{\sigma}}=t_0\hat{N}+U\sum_i\hat{n}_{i\uparrow}\hat{n}_{i\downarrow},
 \label{eq:single_site}
\end{equation}
where $t_{ii}=t_0$ is regarded as constant for a translationally invariant system, as is $\hat{N}=N$ (in general). The two-site terms are
\begin{equation}
\begin{split}
I_2 &= \sum_{ij}\!{'} \bigg\{ V_{iijj} \hat c_{i\uparrow}^{\dagger}\, \hat c_{i\downarrow}^{\dagger}\, \hat c_{j\downarrow}\, \hat c_{j\uparrow} - V_{ijji} \left( {\bf \hat S}_i \cdot {\bf \hat S}_j
+ \frac{1}{4} \hat n_i \hat n_j \right)\\
&+ \frac{1}{2} V_{ijij}\, \hat n_i \hat n_j \bigg\}.
\end{split}
\label{eq:two_site}
\end{equation}
Now, if the single-particle basis is selected as real, we have $V_{iiij}=V_{iiki}=V_{ijji}=V_{ijjj}-V_{iiij}=K_{ij}$, $V_{ijij}\equiv V_{ij}$, and $J_{ij}\equiv V_{ijji}$.

In effect, putting $I_1$ and $I_2$ together, we obtain
\begin{equation}
\begin{split}
\mathcal{\hat H} = &\, t_0 \sum_{i\sigma}  \hat n_{i\sigma} + \sum_{ij\sigma}\!^{'} t_{ij}\, \hat c_{i\sigma}^{\dagger}\, \hat c_{j\sigma} + U \sum_{i}  \hat n_{i\uparrow}\,  \hat n_{i\downarrow}\\
+ &\frac{1}{2} \sum_{ij}\!^{'} \left(V_{ij} - \frac{1}{2} J_{ij}\right) \hat n_i\, \hat n_j - \sum_{ij}\!^{'} J_{ij}\, {\bf \hat S}_i \cdot {\bf \hat S}_j\\
+ &\sum_{ij}\!^{'} J_{ij}\,\hat c_{i\uparrow}^{\dagger}\, \hat c_{i\downarrow}^{\dagger}\, \hat c_{j\downarrow}\, \hat c_{j\uparrow}\\
+ &\frac{1}{2} \sum_{ij\sigma}\!^{'} K_{ij} \left( \hat n_{i\bar{\sigma}} +  \hat n_{j\bar{\sigma}} \right) \left( \hat c_{i\sigma}^{\dagger}\, \hat c_{j\sigma} + \hat c_{j\sigma}^{\dagger}\, \hat c_{i\sigma} \right).
\end{split}
\label{eq:general_final}
\end{equation}
The first term can be dropped as it defines the reference point for the chemical potential ($t_0=0$), the second and the third define the Hubbard model (primed summation means that $i\neq j$) the fourth term defines the direct intersite Coulomb term (in the text we put $V\equiv V_{\langle ij\rangle}-J_{\langle ij\rangle}/2$), the fifth represents the Heisenberg exchange, the sixth corresponds to the local pair hopping (neglected in our previous work \cite{Spalek2017}, as we considered there only the large-$U$ limit), and the last is the correlated hopping term with $K\equiv 2K_{\langle ij\rangle}$. Those terms can be grouped into the form presented here in the starting Hamiltonian (\ref{eq:H_start}) by changing the summation indices properly $i\leftrightarrows j$ in consecutive terms. However, it should be noted that the Heisenberg exchange term presented here should not by associated with the superexchange interaction term $\sim J$ which appears in our starting Hamiltonian (\ref{eq:H_start}). The latter 
originates from virtual hopping processes through the 2$p\sigma$ states from the Cu-O plane\cite{Spalek2017}. The purpose of this detailed derivation is to show that the Hamiltonian (\ref{eq:H_start}) contains all important two-site terms for the correlated electrons within a single-band. Therefore, it should be tested extensively in the 
context of high-temperature superconductivity and associated with it other forms of orderings\cite{Keimer2015}. Additionally, we have listed here the assumptions required to derive it. For $3d$ electrons in $d_{x^2-y^2}$ state one really needs the real-valued representation of the wave function. Finally, the Hamiltonian (\ref{eq:general_final}) in the strong correlation limit can be transformed directly into an extended t-J model \cite{Spalek1981}.


\begin{thebibliography}{99}

\bibitem{Anderson1997}
P. W. Anderson in \textit{Frontiers and Borderlines in Many-Particle Physics.} Editors: R. A. Broglia and J. R. Schrieffer, North-Holland, Amsterdam 1988, pp. 1-47.

\bibitem{Dagotto1990}
E. Dagotto, Rev. Mod. Phys. {\bf 66}, 763 (1994).

\bibitem{Zhang1988}
F. C. Zhang, and T. M. Rice, Phys. Rev. B {\bf 37}, 3759 (1988).

\bibitem{Ogata2008}
M. Ogata and H. Fukuyama, Rep. Prog. Phys. {\bf 71}, 036501 (2008).

\bibitem{Anderson1988}
J. Spa\l ek, Phys. Rev. B {\bf 37}, 533 (1988).

\bibitem{Anderson2004}
P. W. Anderson, P. A. Lee, M. Randeria, T. M. Rice, N. Trivedi, and F. C. Zhang, J. Phys.: Condens Matter. 16 R755-R769 (2004).

\bibitem{Randeria2012}
R. Randeria, R. Sensarma, and N. Trivedi in \textit{Projected Wavefunctions and High-T$_c$ Superconductivity in Doped Mott Insulators}, in: \textit{Strongly Correlated Systems: Theoretical Methods}, eds. A. Avella and F. Macini (Springer Verlag, Berlin, 2012) Chapter 2, pp. 29-64.

\bibitem{Zhang1988_2}
F. C. Zhang, C. Gros, T. M. Rice, and H. A. Shiba, Supercond Sci Technol. {\bf 1}, 36 (1988).

\bibitem{Edegger2007}
B. Edegger, V. N. Muthukumar, and C. Gros, Adv. Phys. {\bf 56}, 927 (2007).


\bibitem{Eichenberger}
D. Eihenberger and D. Baeriswyl, Phys. Rev. B {\bf 76}, 180504 (2007).

\bibitem{Kaczmarczyk2013} 
J. Kaczmarczyk, J. Spa{\l}ek, T. Schickling, and J. B\"unemann, Phys. Rev. B {\bf 88}, 115127 (2013).

\bibitem{Spalek2017}
J. Spa\l ek, M. Zegrodnik, and J. Kaczmarczyk, Phys. Rev. B {\bf 95}, 024506 (2017).

\bibitem{Zegrodnik2017}
M. Zegrodnik, J. Spa\l ek Phys. Rev. B 95, 024507 (2017).

\bibitem{Wojtowicz2001}
J. Wojtowicz and R. Lema\'nski, Phys. Rev. B {\bf 64} 233103 (2001).

\bibitem{Shavika}
A. M. Shvaika, Phys. Rev. B {\bf 67} 075101 (2003).

\bibitem{Stauber}
T. Stauber and F. Guinea, Phys. Rev. B {\bf 69} 035301 (2004).

\bibitem{Hubsch2006}
A. H\"ubsch, J. C. Lin, J. Pan and D. L. Cox, Phys. Rev. Lett. {\bf 96} 196401 (2006).

\bibitem{Vitoriano2009}
C. Vitoriano and M. D. Coutinho-Filho, Phys. Rev. Lett. {\bf 102} 146404 (2009).

\bibitem{Gorski2011}
G. G\'orski and J. Mizia, Phys. Rev. B {\bf 83} 064410 (2011).

\bibitem{Hirsh}
J. E. Hirsch and F. Marsiglio, Phys. Rev. B {\bf 39} 11515 (1989).

\bibitem{Marsiglio}
F. Marsiglio and J. E. Hirsch, Phys. Rev. B {\bf} 41 6435 (1990).

\bibitem{Micnas1990}
R. Micnas, J. Ranninger and S. Robaszkiewicz, Rev. Mod. Phys. {\bf} 62 113 (1990).


\bibitem{Wysokinski2017}
M. M. Wysoki\'nski, J. Kaczmarczyk, J. Phys. Condens. Matter {\bf 29} 085604 (2017).  
  
  
\bibitem{Deutscher2005}
G. Deutscher, A. F. Santander-Syro, and N. Bontempt, Phys. Rev. B {\bf 72}, 092504 (2005).

\bibitem{Carbonne2006}
F. Carbonne \textit{et al.}, Phys. Rev. B {\bf 74}, 064510 (2006).

\bibitem{Molegraaf2002}
H. J. A. Molegraaf, Science {\bf 295}, 2239 (2002).

\bibitem{Gianetti2011}
C. Gianneti \textit{et al.}, Nat. Comm. {\bf 2} (353), 1 (2011).


\bibitem{Zhou2003}
X. J. Zhou \textit{et al.}, Nature {\bf 423}, 398 (2003).

\bibitem{Kordyuk2005}
A. A. Kordyuk, S. V. Borisenko, A. Koitzsch, J. Fink, M. Knupfer, and Berger, Phys. Rev. B {\bf 71}, 214513 (2005).

\bibitem{Borisenko2006}
S. V. Borisenko et al., Phys. Rev. Lett. {\bf 96}, 117004 (2006).

\bibitem{Hashimoto2008}
M. Hashimoto et al., Phys. Rev. B {\bf 77}, 094516 (2008).

\bibitem{Abram2016}
Abram M, Zegrodnik M, and Spa\l ek J 2016 arxiv:1607.05399.

\bibitem{Laughlin2014}
R. B. Laughlin, Phys. Rev. B {\bf 89}, 035134 (2014).


\bibitem{Bunemann2012} 
J. B\"unemann, T. Schickling, and F. Gebhard, Europhys. Lett. {\bf 98}, 27006 (2012).

\bibitem{Kaczmarczyk2014}
J. Kaczmarczyk, J. B\"unemann, and J. Spa{\l}ek, New J. Phys. {\bf 16}, 073018 (2014).


\bibitem{Wysokinski2015}
Wysoki\ifmmode~\acute{n}\else \'{n}\fi{}ski M~M, Kaczmarczyk J and Spa\l{}ek J, Phys. Rev. B {\bf 92} 125135 (2015).
  
\bibitem{Wysokinski2016}  
M. M. Wysoki\'nski, J. Kaczmarczyk, and J. Spa\l ek, Phys. Rev. B {\bf 94}, 024517 (2016).

\bibitem{Kaczmarczyk2016n}
J. Kaczmarczyk, T. Schickling, and J. B\"unemann, Phys. Rev. B {\bf 94} 085152 (2016).

\bibitem{zuMunster2016}
K. zu M\"unster and J. B\"unemann, Phys. Rev. B {\bf 94} 045135 (2016).

\bibitem{Hubbard1963}
J. Hubbard, Proc. R. Soc. A {\bf 276} 238 (1963).

\bibitem{Vishik2010}
I. M. Vishik, W. S. Lee, F. Schmit, B. Mortiz, T. Sagawa, S. Uchida, K. Fujita, S. Ishida, C. Zhang, T. P. Devereaux, and Z. X. Shen, Phys. Rev. Lett. {\bf 104}, 207002 (2010).

\bibitem{Keimer2015}
B. Keimer, S. A. Kivelson, M. R. Norman, S. Uchida, and J. Zaanen, Nature (London) {\bf 518}, 179 (2015).

\bibitem{Drachuck2014}
G. Drachuck, E. Razzoli, R. Offer, G. Batalisky, R. S. Dhaka, A. Kanigel, M. Shi, and A. Keren, Phys. Rev. B {\bf 89}, 121119(R) (2014).

\bibitem{Park2008}
S. R. Park, D. J. Song, C. S. Leem, C. Kim, C. Kim, B. J. Kim, and H. Eisaki, Phys. Rev. Lett. {\bf 101}, 117006 (2008).

\bibitem{Armitage2003}
N. P. Armitage, D. H. Lu, C. Lim, A. Damascelli, K. M. Shen, F. Ronning, D. L. Feng, P. Bogdanov, X. J. Zhou, W. L. Yang, Z. Hussain, P. K. Mang, N. Kaneko, M. Greven, Y. Onose, Y. Taguchi, Y. Tokura, and Z.-X.Shen, Phys. Rev. B {\bf 68}, 064517 (2003).

\bibitem{Anderson1973}
P. W. Anderson, Mater. Res. Bull., {\bf 153} (1973).

\bibitem{Fazekas1974}
P. Fazekas and P. W. Anderson, Phil. Mag. {\bf 30}, 23 (1974).

\bibitem{Randeria2016}
M. Randeria, in PWA {\bf 90} A liftime of Emergence (World Scientific, New Jersey, 2016) edited by P. Chandra et al. pp. 113-126.

\bibitem{Kaminski2007}
T. Kondo et al., Phys. Rev. Lett. {\bf 98}, 267004 (2007).

\bibitem{Yoshida2012}
T. Yoshida et al., J. Phys. Soc. Jpn. {\bf 81}, 011006 (2012).

\bibitem{Vishik2012}
I. M. Vishik et al., PNAS {\bf 109}, 18332 (2012).

\bibitem{Biborski2016}
A. Biborski, A. K\k{a}dzielawa, A. Gorczyca-Goraj, E. Zipper, M. M. Ma\'ska, and J. Spa\l ek, Sci. Rep. 6, 29887 (2016).

\bibitem{Spalek1981}
J. Spa\l ek, A. M. Ole\'s, and K. A. Chao, Phys. Stat. Solidi (b) {\bf 108}, 329 (1981).


\end{thebibliography}
\end{document}